\documentclass{ws-procs9x6-cpt16}
\begin{document}

\newcommand{\refeq}[1]{(\ref{#1})}
\def\etal {{\it et al.}}

\def\al{\alpha}
\def\be{\beta}
\def\ga{\gamma}
\def\Om{\Omega}
\def\cF{{\cal F}}
\def\fr#1#2{{{#1} \over {#2}}}
\def\sb{\overline{s}{}}
\def\afb{\alpha(\overline{a}_{\rm eff})}
\def\epd{s^{(d)}}
\def\mn{{\mu\nu}}
\def\Fd{\cF^w(d)}
\def\G{G_N}

\title{Gravitational Searches for Lorentz Violation with\\
Matter and Astrophysics}

\author{Jay D.\ Tasson}

\address{Physics Department, St. Olaf College,
Northfield, MN 55057, USA}

\begin{abstract}
This contribution to the CPT'16 proceedings
summarizes recent tests of Lorentz violation in the pure-gravity sector
with cosmic rays
and reviews recent progress in matter-gravity couplings.
\end{abstract}

\bodymatter

\section{Introduction}
Lorentz violation\cite{rev}
as a signal of new physics at the Planck scale\cite{ks}
is actively sought in a wide variety of tests\cite{data}
within the general test framework of the gravitational Standard-Model Extension (SME).\cite{SME}
A subset of these tests involve gravitational physics,
which can probe Lorentz violation in the pure-gravity sector
associated with both minimal\cite{lvpn} and nonminimal\cite{cer,short,ligo} operators,
as well as Lorentz violation in matter-gravity couplings.\cite{lvgap,akjt}
While Lorentz violation in gravity has been sought in a number of systems,\cite{data}
we focus here on recent tests exploring gravitational \v Cerenkov radiation
in Sec.\ \ref{cos},
and review the status of searches for the $\afb_\mu$ coefficient
in the matter sector in Sec.\ \ref{matter},
including work with superconducting gravimeters.\cite{fgt}

\section{Gravitational \v Cerenkov radiation}
\label{cos}
The \v Cerenkov radiation of photons by charged particles moving faster than the phase speed of light
in ponderable media
is a well-known phenomenon in Nature.
If the analogous situation of particles exceeding the phase speed of gravity
were to occur,
as might be the case in General Relativity (GR) in the presence of certain media,\cite{grcer}
the gravitational \v Cerenkov radiation of gravitons would be expected.\cite{sz71}
In the presence of suitable coefficients for Lorentz violation in the SME,
both electrodynamic\cite{emcer} and gravitational\cite{cer} \v Cerenkov radiation become possible in vacuum.
In this section,
we review the tight constraints that have been achieved
by considering cosmic rays
and mention other possible implications of vacuum gravitational \v Cerenkov radiation.\cite{cer}

As with all SME searches for Lorentz violation,
our analysis begins with the expansion about GR and the Standard Model provided by the SME action.
Here we consider the linearized pure-gravity sector,
which has now been written explicitly for operators of arbitrary mass dimension $d$
that preserve the usual gauge invariance of GR.\cite{ligo}
Except where noted,
we assume that the other sectors of the theory are conventional.
Exploration of the dispersion relation 
generated by this action reveals that a class of coefficients for Lorentz violation
manifest as a momentum-dependent metric perturbation
that generates a momentum-dependent effective index of refraction for gravity.
With an appropriate sign for the coefficients for Lorentz violation,
this index is greater than 1
and particles may exceed the speed of gravity and radiate gravitons.
Hence each observation of a high-energy particle
can place a one-sided constraint on a combination of these coefficients.

To obtain constraints,
we perform a calculation of the rate of graviton emission
that parallels standard methods
assuming for simplicity and definiteness
that only coefficients at one arbitrary dimension $d$ are nonzero.
The calculation is provided for photons,
massive scalars,
and fermions,
the only differences in the three cases being 
the details of the matrix element for the decay
and the form of a dimensionless function of $d$ in the results. 
The rate of power loss can then be integrated
to generate a relation between the time of flight $t$ for
a candidate graviton-radiating particle,
the energy of the particle at the beginning of its trip $E_i$,
and the observed energy at the end of its trip $E_f$.
This relation takes the form
\begin{equation}
t = \fr {\Fd} {\G (\epd)^2}
\left(\fr{1}{E_f^{2d-5}}-\fr{1}{E_i^{2d-5}}\right),
\label{tflight}
\end{equation}
where $G_N$ is Newton's constant,
$\Fd$ is a species $w$ dependent function of $d$,
and $\epd$ is a combination of coefficients for Lorentz violation
at dimension $d$ that depends in general on the direction of travel for the particle.
This result is distinguished from earlier work 
on the subject of gravitational \v Cerenkov radiation\cite{mncer}
by the connection to the field-theoretic framework of the SME,
the consideration of anisotropic effects,
the exploration of arbitrary dimension $d$,
and the treatment of photons and fermions.

Conservative constraints can be placed using Eq.\ \eqref{tflight}
by setting $1/E_i^{2d-5} = 0$,
solving for $\epd$,
\begin{equation}
\hskip -20pt
\epd(\hat p) \equiv (\sb^{(d)})^{\mn\al_1 \ldots \al_{d-4}}
\hat p_\mu \hat p_\nu 
\hat p_{\al_1} \ldots \hat p_{\al_{d-4}}
< \sqrt{\fr{\Fd}{\G E_f^{2d-5}L}},
\label{bound}
\end{equation}
and using suitable data on cosmic ray observations.\cite{cosmic}
Here $L$ is the travel distance.
Given the dependence on $E_f$ in Eq.\ \eqref{bound},
the highest energy events yield the tightest bounds.
The highest energy cosmic rays are believed to be nuclei,
and continuing toward conservative constraints,
we assume that the gravitons are radiated by a partonic fermion in an iron nucleus
carrying 10\% of observed energy $E_\oplus$,
which leads to $E_f = E_\oplus/560$.
A consideration of the likely origin of these particles
leads to a conservative estimate of $L=10$~Mpc.
We then use the available data on cosmic ray energies
and direction of origin\cite{cosmic} to place constraints on six models.
Three of the models are constructed as the isotropic limit
at $d=4,6,8$ respectively.
In each of these models we place a one-sided limit on the one isotropic coefficient involved
at the level of $10^{-14}$, 
$10^{-31}$ GeV$^{-2}$,
and $10^{-48}$ GeV$^{-4}$ respectively.
The other three models involve two-sided constraints on the anisotropic coefficients at each $d$.
At $d=4$ we place eight constraints at the $10^{-13}$ level,
while at $d=6$ we constrain 24 coefficients at the level of $10^{-29}$ GeV$^{-2}$,
and at $d=8$ we constrain 48 coefficients at the $10^{-45}$ GeV$^{-4}$ level.

The paper concludes by discussing some  ways in which our work might be extended.
Topics considered include the role of the matter sector,
the impact of gravitational \v Cerenkov radiation by photons on cosmological models,
and gravitons emitting electromagnetic \v Cerenkov radiation.

\section{Matter-gravity couplings}
\label{matter} 
In Ref.\ \refcite{lvgap} the phenomenology
of matter-gravity couplings was developed
with a focus on spin-independent coefficients,
particularly the countershaded\cite{akjt} $\afb_\mu$ coefficients.
As of the CPT'13 meeting,\cite{jtcpt13}
constraints on $\afb_\mu$ had been placed using the following systems:\cite{data,jtcpt13}
precession of the perihelion of Mercury\cite{lvgap} and Earth,\cite{lvgap}
torsion pendula,\cite{akjt}
a torsion strip balance,\cite{speake}
atom interferometry,\cite{hohensee}
and co-magetometry.\cite{gmag}
This work resulted in a number of measurements of the time component
reaching the level of $10^{-11}$ GeV on both the neutron and the proton plus electron coefficients.
For the spatial components,
two combinations of the nine coefficients were constrained at the level of $10^{-6}$ GeV,
and four combinations were weakly constrained at the $10^{-1}$ GeV level.
Note that coverage is sufficient to span the space that is accessible
without charged-matter experiments.

Since CPT'13, $\afb_\mu$ (as well as $\sb_\mn$) 
has been considered in an analysis of planetary ephemerides.\cite{hees}
This work considerably extends the level of the independent constraints on $\afb_J$ coefficients 
to $10^{-5}$ GeV to $10^{-3}$ GeV,
and the analysis of gravimeter experiments extends the maximum reach for these coefficients
even further.\cite{fgt}
The
consideration of bound kinetic energy in equivalence-principle tests\cite{hmw}
has also been used to further separate the $\afb_T$ coefficients and other matter-sector
coefficients.
Though the $\afb_\mu$ coefficient space accessible with ordinary matter has now been
covered more uniformly with initial constraints,
opportunities for further improvements with currently available methods remain.\cite{lvgap,jtpuls}

\end{document}